\documentstyle[epsf,aas2pp4]{article}

\newcommand{\etal}{{et al.}\hspace{1mm}}
\newcommand{\nh}{N_{\rm HI}}
\newcommand{\db}{\delta_b}

\begin{document}


\title{The $b$ Distribution of the Ly$\alpha$ Forest: Probing
Cosmology and the Intergalactic Medium}

\author{Greg L. Bryan\altaffilmark{1}}
\affil{Department of Physics, Massachusetts Institute of Technology,
Cambridge, MA 02139}

\and

\author{Marie E. Machacek}
\affil{Department of Physics, Northeastern University, Boston, MA 02115}

\altaffiltext{1}{Hubble Fellow}


\begin{abstract}

We investigate a method to determine the temperature-density relation
of the intergalactic medium (IGM) at $z \sim 2-4$ using quasar
absorption line systems.  Using a simple model combined with numerical
simulations we show that there is a lower cutoff in the distribution
of column density ($\nh$) and line width ($b$ parameter).  The
location of this cutoff can be used to determine the
temperature-density relation (under certain conditions).  We describe
and test an algorithm to do this.
The method works as long as the amplitude of fluctuations on these
scales ($\sim 100$ kpc) is sufficiently large.  Models with less
power can mimic higher temperatures.  A preliminary application is made
to data from two quasar lines-of-sight, and we determine an upper
limit to the temperature of the IGM.  Finally, we examine the full
distribution of $b$-parameters and show that this is completely
specified by just two parameters: the temperature of the gas and the
amplitude of the power spectrum.  Using the temperature upper limit
measured with the $\nh-b$ cutoff method, we derive an upper limit to
the amplitude of the power spectrum.

\end{abstract}


\twocolumn


\section{Introduction}

It has become clear that observations of absorption lines in the
spectra of high-redshift quasars can give us valuable information
about the nature and distribution of the intergalactic medium.  Early
theoretical work (\cite{dor77}; \cite{ree86}; \cite{bon88};
\cite{mcg90}; \cite{bi92}), supplemented with numerical simulations
(\cite{cen94}; \cite{pet95}; \cite{zha95}; \cite{her96}) showed
convincingly that absorption lines at $z \sim 3$ with column densities
less than about $10^{16}$ cm$^{-2}$ arise primarily from a network of
relatively low density filaments and sheets that naturally form out
of hierarchical primordial perturbations.

Having established the link between cosmology and the Ly$\alpha$
forest, subsequent work has focused on two related areas: improving
our understanding of the physical conditions of the IGM (at these
redshifts), and using the forest to constrain cosmological parameters.
This includes using the power spectrum of the flux distribution
(\cite{cro98}; \cite{cro99}), the slope of the column density
distribution (\cite{hui97}; \cite{gne98}; \cite{mac99}), and inverting
the flux-density relation (\cite{nus99}).

Although early simulations seemed to show that all models were in
agreement with observations, recently it was shown (\cite{the98};
\cite{bry99}), that the width of the absorbers, commonly quantified by
the $b$ parameter of a Voigt-profile, had been over-predicted in most
previous work.  This left a discrepancy between the canonical model
and the observations.

A number of ways to resolve this have been suggested, mostly revolving
around ways to increase the temperature of the gas, and hence the
width of the lines.  The low density gas in the IGM is very close to
photoionization equilibrium with a background radiation field, usually
assumed to be from quasars.  Its temperature is determined by a
competition between adiabatic cooling and photo-ionization heating
(\cite{hui97b}).  An increase in the density will result in more
photo-ionization heating and hence higher temperatures.  In Theuns
\etal (1999)\markcite{the99}, it was shown that increasing $\Omega_b$,
the ratio of the baryon density to the critical density, could widen
the lines.  However, even after doubling the baryon density to the
edge of that permitted by primordial nucleosynthesis, they still found
some disagreement.  Along similar lines, delaying helium reionization
to $z \sim 3-4$ (\cite{hae97}) can provide a small boost in the
temperature.

In part driven by this discrepancy, there have been some recent
suggestions on other ways to increase the temperature of the IGM.  The
first is a suggestion of Compton heating from a hard X-ray background
(\cite{mad99}).  The second stems from the observation that the
commonly adopted optically-thin limit for photo-ionization heating
(particularly for helium) may result in a substantial underestimate of
the gas temperature (\cite{abe99}).  A third, which we will not
examine in detail in this paper, is provided by photoelectric heating
from dust grains (\cite{nat99}).  Each of these could, in principle,
provide the factor of two increase in the temperature required.

However, since the width of the lines is not just due to temperature
but also comes from the velocity structure (both peculiar and Hubble
velocities) along the line-of-sight, it seems likely that other
parameters also play a role.  In an elegant paper based on linear
perturbation theory, Hui \& Rutledge (1998)\markcite{hui98} argued
that the width should depend inversely on the amplitude of the
primordial density fluctuations.  

In this paper we show that there exists a way to indirectly measure
the temperature of the IGM.  The method is based on a lower cutoff in
the $\nh-b$ distribution (first noted by \cite{zha97a}).  The position
and slope of this line is a reflection of the density-temperature
relation of the IGM.  A simple model for this is presented in
section~\ref{sec:theory} and extensive tests using numerical
simulations are described in section~\ref{sec:nhb}.  We develop
a simple but robust statistic to find the location and slope of the
cutoff in the $\nh-b$ plane.

However, we also demonstrate that this method breaks down if the
amplitude of the density fluctuations is too low.  Again, we present a
simple explanation for why this occurs and show directly with
simulations that it can mimic the effect of higher-temperature gas.
This means that the density-temperature relation derived in this way
must be treated as an upper limit (until the power spectrum can be
fixed by other means).

Switching from the cutoff in the $\nh-b$ distribution to the full
distribution of $b$ parameters, we show in section~\ref{sec:median}
that the entire distribution is controlled by the same two parameters
described above: the temperature of the gas and the amplitude of the
primordial fluctuations.  In fact, in this case these two variables are
completely degenerate and form a single parameter.

In the last section (section~\ref{sec:observations}), we apply our
tests to previously published observations of two quasars, and derive
a temperature density relation.



\section{Theory}
\label{sec:theory}

First, we quickly review the calculation of the line profile; more
complete discussions can be found elsewhere (\cite{hui97};
\cite{zha97b}).  The optical depth at a given (observed) frequency
$\nu_0$ can be calculated with:
\begin{equation}
\tau(\nu_0) = \int_{x_A}^{x_B} n_{\rm HI} \sigma_\alpha \frac{dx}{1+z},
\end{equation}
where $x$ is the comoving radial coordinate along the line-of-sight,
and $n_{\rm HI}$ is the neutral hydrogen density at this point (with
redshift $z$).  The Ly$\alpha$ cross-section $\sigma_\alpha$ is a
function of the frequency of the photon with respect to the rest frame
of the gas at position $x$:
\begin{equation}
\nu = \nu_0 \left(1+z\right) \left(1 + \frac{v_{pec}}{c}\right).
\end{equation}
Here, $v_{pec}$ is the peculiar velocity and $z$ is the redshift
due to the Hubble expansion only.  This can be rewritten in terms
of the velocity:
\begin{equation}
\label{eq:u}
u = \frac{H}{1+z}(x - x_0) + v_{pec}(x),
\end{equation}
where we are expanding around the point $x_0$, and $H$ is the Hubble
``constant'' at this redshift.  The expression is valid as long as
$u/c$ is much smaller than 1.  In this case, the optical depth can be
written as:
\begin{equation}
\label{eq:tau}
\tau(u_0) = \sum \int_{u_A}^{u_B} \frac{n_{\rm HI}}{1 + z} 
            {\left| \frac{du}{dx} \right|}^{-1} \sigma_\alpha du.
\end{equation}
The summation sign arises because equation~(\ref{eq:u}) can be
multi-valued.  The cross-section, assuming that Doppler broadening
dominates over natural or collisional line-broadening (accurate for
column densities less than $10^{17}$ cm$^{-2}$), is given by:
\begin{equation}
\sigma_\alpha = \sigma_{\alpha,0} \frac{c}{b\sqrt{\pi}} 
                e^{-(u-u_0)^2/b^2}.
\end{equation}
Where we have used the following standard definitions: $b = (2
k_b T / m_p)^{1/2}$, $k_b$ is the Boltzmann constant, $m_p$ is the proton
mass and $\sigma_{\alpha,0}$ is the Ly$\alpha$ line-center
cross-section. 

\subsection{Measuring the temperature of the IGM}
\label{sec:eos}

As outlined in the introduction, a central question is how to
determine the temperature-density relation of the gas.  We also
loosely refer to this as the equation-of-state.  It is quantified as: 
\begin{equation}
\label{eq:eos}
T = T_0 (1 + \delta_b)^{\gamma - 1}.
\end{equation}
Here, $(1+\delta_b) = \rho_b / (\Omega_b \bar{\rho})$, where
$\bar{\rho}$ is the mean density (both baryonic and dark).  For gas
which is primarily heated by UV photoionization, $\gamma$ is expected
to vary from 1 immediately after reionization, to a limiting value of
about 1.5 (\cite{hui97b}).  Similarly, $T_0$ is expected to evolve as
$T_0 \sim (1+z)^{1.7}$.  This information is encoded in the absorption
lines, and here we describe a way to indirectly measure, or
at least constrain, the equation of state.

The width of a given line is the result of a convolution involving
both the temperature and velocity distributions of the gas along the
line of sight.  However, for low column-density lines ($ \le
10^{15}$ cm$^{-2}$) both the density and temperature are slowly varying
functions of position (\cite{bry99}).  Therefore, it makes sense to
distinguish two sources of the total line-width:
\begin{itemize}
 \item $b_T$, the thermal Doppler broadening, which is a measure of the
optical-depth weighted temperature of the gas, and
 \item $b_{vel}$, the bulk velocity broadening, which in turn has
two components: the peculiar velocity and the Hubble velocity.
\end{itemize}
The total width comes from adding these two in quadrature:
$b_{total} = (b_T^2 + b_{vel}^2)^{1/2}$.

Our method is based on two assumptions.  First, that the
column density of a line is proportional to the density of the gas,
so a measurement of $\nh$ can be converted to $(1+\db)$.
The second assumption is that there are at least some 
lines (at a given $\nh$) for which $b_{vel}$ is
significantly smaller than $b_T$ so that $b_{total} \approx b_T$.
If this is true, then there will be a minimum in the
$\nh-b$ distribution given by:
\begin{eqnarray}
\label{eq:bmin}
b_{\rm min} \approx b_T & = & (2 k_b T / m_p)^{1/2} \\
        & = & (2 k_b T_0 (1+\db)^{\gamma-1} / m_p)^{1/2},
                 \nonumber
\end{eqnarray}
where $1 + \delta_b$ is the mean overdensity
of a line with column density $\nh$.  In fact, this argument was
first suggested by Zhang \etal (1997)\markcite{zha97a}.

To make this a little more concrete, we can generate a toy model for
this minimum based on a number of assumptions: (1) photoionization
equilibrium holds, (2) when computing column densities, peculiar
velocities can be ignored, (3) all systems have the same comoving
length, and (4) at any column density, there exist some absorbers with
$b_{vel} = 0$.  None of these conditions hold exactly, however
numerical simulations show that --- for at least some models --- 
they are not unreasonable approximations (\cite{bry99}).  In fact,
as we will show, it is the last assumption that will present the
most difficulties.

From the first assumption, it is relatively straightforward
(e.g. \cite{zha97b}) to show that the neutral hydrogen density is
given by:
\begin{equation}
\label{eq:nh}
n_{\rm HI} = 1.2 \times 10^{-16} \left(1+\delta_b\right)^{2}
         \left(\frac{\Omega_b h^2}{0.02} \right)^2
         \frac{\left(1+z\right)^6}{\Gamma_{{\rm HI,}-12}}
	 {T_4}^{-0.7} 
         \hbox{cm}^{-3},
\end{equation}
where $T_4 = T/10^4 K$ and $\Gamma_{{\rm HI,}-12}$ is the hydrogen
photoionization rate in units of $10^{-12}$ s$^{-1}$.  The next two
assumptions provide a relation between this and the column density,
which we will take to be: $\nh = l n_{\rm HI}$ where $l = 125$
$(4/(1+z))$ kpc, so that:
\begin{equation}
\label{eq:nh_delta}
\nh = 1.9 \times 10^{13} F
           (1+\delta_b)^2 T_4^{-0.7} 
           \hbox{cm}^{-2}.
\end{equation}
For notational ease, we have taken the cosmological and
photo-ionization factors into $F$ which is defined as
\begin{equation}
\label{eq:f}
F = \left( \frac{\Omega_b h^2}{0.02} \right)^2
     \left( \frac{1+z}{4} \right)^5 \Gamma_{\rm HI,-12}^{-1}.
\end{equation}
In fact, the exact value of $l$ has been selected to give a 
good fit to the data described below, however it is quite 
compatible with the width of filaments seen in simulations.

Finally, we use
the fourth assumption along with equation~(\ref{eq:eos}) 
to derive an expression for the minimum column
density as a function of temperature:
\begin{equation}
N_{HI,min} = 1.9 \times 10^{13} F 
             T_4^{2/(\gamma-1) - 0.7}
             T_{0,4}^{-2/(\gamma-1)}
             \hbox{cm}^{-2},
\end{equation}
where $T_{0,4}$ is $T_0/10^4$ K.
Using equation~(\ref{eq:bmin}) this can be recast 
entirely in terms of observable quantities
($N_{HI,min}, b_{min}$) and parameters of the equation of state
($T_0$, $\gamma$):
\begin{equation}
\label{eq:b_nh_eos}
b_{min} = 13 \hbox{km/s} 
          {\left(\frac{N_{HI,min}}{1.9 \times 10^{13}
          \hbox{cm}^{-2} F 
                       } 
            \right)}^\frac{\gamma-1}{5.4-1.4\gamma} 
          T_{0,4}^{1/(2.7-0.7\gamma)}.
\end{equation}
This expression shows that the minimum in the $\nh-b$ distribution
should take the form of a power law (since the equation of state is
assumed to be a power law).  It gives a way to determine the
parameters of the density-temperature relation in
equation~(\ref{eq:eos}) from a measurement of the intercept and slope
of the minimum line in the $\nh-b$ plane.

\subsection{The cosmology-$b$ connection}

What else, besides temperature, influences the width of the absorbers?
The most significant cosmological parameter turns out to be the 
amplitude of the primordial density fluctuations on the 
scales giving rise to the forest.  The most
convincing demonstration of this comes from numerical simulation;
however, a simple plausibility argument can be made as follows.
\footnote{Another calculation along these
lines, but for a random Gaussian field instead of a single
perturbation, can be found in Hui \& Rutledge (1998), where they also
derived the expected shape of the distribution of $b$
parameters.\markcite{hui98}} 

We begin with a sinusoidal perturbation of comoving wavenumber $k$
which perturbs a fluid element's Lagrangian position $q$:
\begin{equation}
\label{eq:sin}
x = q - D_+ A \sin(kq)/k,
\end{equation}
where $A$ is the initial amplitude and $D_+$ describes the evolution
of a growing mode for the given cosmology ($D_+ \propto (1+z)^{-1}$ in
an Einstein-deSitter universe).  This equation usually begins a
discussion of the Zel'dovich approximation but here we will need to
assume that $A$ is small compared to unity.

The peculiar velocity of the fluid element is given by
\begin{equation}
v_{pec} = a \dot{x} = \dot{a} f D_+ A \sin(kq)/k.
\end{equation}
We adopt the common notation $f = a \dot{D}_+ / \dot{a} D_+$, where
$a=(1+z)^{-1}$ is the scale factor (\cite{pee93}).  For gas in
photo-ionization equilibrium, the neutral hydrogen density is related
to the gas density by $n_{\rm HI} \sim \rho^{1.7}$.  The density
produced by this perturbation is given by $\rho \sim \left(1+D_+ A
\cos(kq)\right)$ so,
\begin{equation}
n_{\rm HI} \propto \left( 1 +       D_+ A \cos(kq) \right)^{-1.7}
       \propto \left( 1 - 1.7 D_+ A \cos(kq) \right).
\end{equation}
For brevity, we have dropped the coefficients to this expression since
they contribute only to the overall normalization of the optical
depth, not the structure of the line.

This is the density in physical space.  In order to compute the
redshift-space density and hence the optical depth distribution via
equation~(\ref{eq:tau}), we need the Jacobian,
\begin{equation}
\left|\frac{du}{dx}\right|^{-1} = \frac{1}{\dot{a}} 
       \left(1 + f D_+ A \cos(kq)\right).
\end{equation}
Using the previous two expressions, the full expression for the optical
depth distribution is given by:
\begin{equation}
\label{eq:tau_linear}
\tau \propto \int \left(1 - (1.7-f)D_+ A \cos(ku/\dot{a}) \right) 
	\sigma_\alpha du,
\end{equation}
where we have employed equation~(\ref{eq:u}) (to first order in $A$) to
write an expression only in terms of $u$.

While this is helpful, there is still a convolution with the Doppler
width to contend with.  The general expression is quite complicated;
it is more helpful to recognize that the result of the convolution
will be quite close to a Gaussian with width $(b_T^2 +
b_{vel}^2)^{1/2}$ (see Bryan \etal (1999) for an explicit
demonstration of this).  The first term is the Doppler broadening
contribution while the second term comes from the structure of the
line, as given in equation~(\ref{eq:tau_linear}).  Under the
assumption that these two terms are independent, we can ignore the
thermal broadening part and focus simply on the velocity part.
However, the sinusoidal perturbation is not Gaussian, so determining a
$b$ parameter from this is not trivial.  Of course, this is quite true
in the real forest, where line profiles are often not well described
by Voigt-profiles.  We make the correspondence by matching the shapes
of the cosine and Gaussians profiles near the peak of the line, where
the largest contribution to the total optical depth occurs. This is
done by expanding both the cosine term of
equation~(\ref{eq:tau_linear}) and a Gaussian $\exp(-u^2/b^2)$ and
equating the first non-constant term, which is proportional to $u^2$.
Another way to do this would be to start with a Gaussian perturbation
instead of the sinusoidal one in equation~(\ref{eq:sin}).  In either
case, we find that:
\begin{equation}
\label{eq:bvel_full}
b^2_{vel} = \frac{2H^2}{(1.7-f) A D_+ k^2 (1+z)^2}.
\end{equation}

It is interesting to examine this expression in more detail.  For the
redshifts under consideration here, it is useful to approximate the
Hubble velocity as $H = H_0 \Omega_0^{1/2} (1+z)^{3/2}$, which is
accurate as long as $\Omega_0$, the ratio of the total matter density
to that required to close the universe, is not too low (\cite{pee93}).
Similarly, $f \approx \Omega^{0.6} \approx 1$, since $\Omega$
is close to one at this redshifts (again, as long as $\Omega_0$ is not
too low).  By the same reasoning, the growth factor $D_+ \approx
(1+z)^{-1}$.

The wavenumber of the perturbation $k$ is also a factor in this
expression.  While there will be a range of wavelengths, it seems very
reasonable to associate this with the Jeans wavenumber or at least
some fixed fraction of it (e.g. \cite{hui97}; \cite{gnehui97}),
\begin{equation}
\label{eq:kj}
k_j = \sqrt{\frac{12 \pi a^2 G\bar{\rho} \mu}{5 k_b T_0}}.
\end{equation}
In this expression, $\mu$ is the mean mass per particle and is
about $0.6m_p$ for ionized gas.
The average density $\bar{\rho} = 3 H_0^2 \Omega_0 (1+z)^3 / 8 \pi G$
so $k_j \propto H_0^{-1} \Omega_0^{-1/2}$.  (In writing this we have
suppressed a factor of $(1+z)^{-1} T_0^{0.5}$ which is likely to be
quite small since $T_0 \sim (1+z)^{1.7}$).  Therefore, $k \propto
H_0^{-1} \Omega_0^{-1/2} (1+z)$ since it is a comoving wavenumber.

Surprisingly, if we make these assumptions then $H_0$, $\Omega_0$ and
$z$ all cancel, leaving the remarkably simple expression
\begin{equation}
\label{eq:simple}
b_{vel} \propto A^{-0.5}.
\end{equation}
For a more realistic model with a spectrum of perturbations $P(k)$,
the amplitude $A$ will be proportional to the amount of power on the
scales of interest (i.e. $A^2 \propto P(k_j)$).  For a given spectral
shape $b \propto \sigma_8^{-1/2}$.  In fact, approximately this
scaling was found in Machacek \etal (1999).  It fails when the
perturbations become too large or when the thermal width dominates for
a majority of lines; however, it works surprisingly well, even in the
trans-linear regime.  It also explains why the other cosmological
parameters, in particular $\Omega_0$ and the Hubble constant, have so
little effect on the $b$ parameter distribution.  The lack of
a redshift dependence in equation~(\ref{eq:simple}) also helps to
explain why the median distributions (both simulated and observed)
seem to be so constant with redshift, while a 
thermally dominated distribution
would scale as $b \sim T^{1/2} \sim (1+z)^{0.8}$.  We note that this
result differs slightly from that derived in Hui \&
Rutledge\markcite{hui98} (1999), who found $b \propto (D_+
\sigma_8)^{-1/2}$ because they assumed that the smoothing scale would
be constant in redshift space rather than comoving space.

\section{Simulations}
\label{sec:simulations}

In order to examine these effects in more detail, we have performed
numerical simulations of a range of models with various heating rates
and hence various equations of state.  We use a grid-based method
based on the piece-wise parabolic algorithm to model the gas and a
particle-mesh code for the dark matter and gravity.  We follow the
abundances of six species: HI, HII, HeI, HeII, HeIII and e$^-$ by
solving the non-equilibrium evolution equations.  The simulation
method is described in more detail elsewhere (\cite{PPM};
\cite{ann97}).  

Table~\ref{table:simulations} lists the simulations that we analyze in
this paper.  The first column gives the cosmological model --- most are
a form of the currently popular cosmological constant dominated (LCDM)
with $\Omega_0 = 0.4$, $\Omega_\Lambda = \Lambda / 3 H_0^2 = 0.6$,
$\Omega_b = 0.05$, and $h = 0.65$ (the Hubble constant in units of 100
km/s/Mpc).  We also run one other model in order to demonstrate that
the cosmological dependence is well-understood.  This is a flat model
(SCDM) with $\Omega_0 = 1$, $\Omega_b = 0.08$ and $h=0.5$.  The
second column shows the selected normalization of the power spectrum
by giving $\sigma_8$, the linearly extrapolated {\it rms} density
fluctuations in a tophat sphere of 8$h^{-1}$ Mpc.  The next column
gives a measure of small scale fluctuations suggested by Gnedin
(1998)\markcite{gne97}: 
\begin{equation}
\sigma^2_{34} = \int_{0}^{\infty} P(k, z=3) e^{-2k^2/k^2_{34}}
{{k^2 dk} \over {2 \pi^2}}
\end{equation}
where $P(k,z=3)$ is the power spectrum at $z=3$ and $k_{34} = 34
\Omega_0^{1/2} h$ Mpc$^{-1}$.   All power spectra in this paper come
from the analytic fits of Eisenstein \& Hu (1999)\markcite{eis99}.


{
\begin{deluxetable}{ccccccccccc}
\footnotesize
\tablecaption{Simulations analyzed in this paper}
\tablewidth{0pt}
\tablehead{
\colhead{model} & \colhead{$\sigma_8$} & \colhead{$\sigma_{34}$} &
\colhead{L (Mpc)} &
\colhead{$\Gamma_{\rm HeII}/\Gamma_{\rm HeII,HM}$} &
\colhead{X-ray heating?} &
\colhead{$T_{0,4}$} & \colhead{$\gamma$} &
\colhead{$T_{0,4}^\prime$} & \colhead{$\gamma^\prime$} &
\colhead{$b_{\rm med}$ (km/s)} }
\startdata
LCDM & 1.0 & 1.93 & 4.8 & 1.0 & no  & 1.01 & 1.40 & 1.07 & 1.44 & 20.1 \nl
LCDM & 1.0 & 1.93 & 4.8 & 2.0 & no  & 1.30 & 1.39 & 1.35 & 1.43 & 23.0 \nl
LCDM & 1.0 & 1.93 & 4.8 & 4.0 & no  & 1.80 & 1.37 & 1.84 & 1.40 & 27.1 \nl
LCDM & 1.0 & 1.93 & 4.8 & 1.0 & yes & 1.18 & 1.34 & 1.28 & 1.26 & 21.5 \nl
LCDM & 0.8 & 1.54 & 4.8 & 2.0 & no  & 1.52 & 1.34 & 1.36 & 1.42 & 24.6 \nl
LCDM & 0.8 & 1.54 & 4.8 & 1.8 & no  & 1.44 & 1.37 & 1.39 & 1.26 & 23.9 \nl
LCDM & 0.8 & 1.54 & 9.6 & 1.8 & no  & 1.49 & 1.32 & 1.42 & 1.27 & 24.3 \nl
LCDM & 0.6 & 1.16 & 4.8 & 2.0 & no  & 2.31 & 1.22 & 1.44 & 1.32 & 28.9 \nl
SCDM & 0.55 & 1.55 & 4.8 & 2.0 & no  & 1.53 & 1.32 & 1.29 & 1.38 & 24.6 \nl
LCDM$h0.5$ & 0.8 & 1.54 & 4.8 & 2.0 & no & 1.22 & 1.40 & 1.17 & 1.43 & 24.3 \nl
\enddata
\label{table:simulations}
\end{deluxetable}
}


The fourth column indicates $L$ the size of the simulation volume, in
Mpc.  As we demonstrated in a previous paper (\cite{bry99}), there is
some dependence of the $b$-parameter distribution on the box size,
since fluctuations with wavelengths larger than $L$ are not included.
Our canonical box size of 4.8 Mpc is sufficient for a reasonable
prediction, however convergence requires 9.6 Mpc, so we perform one
simulation with this larger size.  All runs use a grid of $128^3$
cells (except the 9.6 Mpc box which uses $256^3$); this provides the
minimum resolution required to accurately resolve the line profiles.

The radiation field is assumed to be spatially constant with the form
given by Haardt \& Madau (1996)\markcite{haa96}, which assumes that
the ionizing photons come from the observed quasar distribution.
However, we modify the HeII photo-heating rate in order to account for
the neglected radiative transfer effects as discussed in Abel \&
Haehnelt (1999)\markcite{abe99}.  Although this is not realistic in
detail it does produce the desired result of heating the IGM.  The
fifth column of Table~\ref{table:simulations} indicates the factor by
which this is increased relative to the original Haardt \& Madau
heating rates.  Abel \& Haehnelt suggested this factor should be $\sim
2-4$.

The next column indicates if the simulation includes Compton heating
due to a hard X-ray background.  We use the heating rate as computed
by Madau \& Efstathiou (1999)\markcite{mad99} who assumed that the
energy density evolved as $U_X(z) = U_X(0) (1+z)^4 \exp(-z^2/z_c^2)$
and included the Klein-Nishina relativistic corrections to the
cross-section, resulting in a heating rate that scales approximately
as $(1+z)^{13/3}$.  We adopt $z_c = 5$.  

The next four columns indicate the equation of state parameters as
given in equation~(\ref{eq:eos}), at $z = 2.7$.  The first set
(without primes) come from fitting the $\nh-b$ minimum and are the
observational estimates, while the second set (with primes) come from
directly fitting 10,000 randomly selected cells in the simulation.
The last column is the median value of the $b$ distribution.

The analysis is carried out by generating artificial spectra along
random lines-of-sight through the computational volume.  These spectra
are then analyzed with an automated Voigt-profile fitting routine
(\cite{zha97a}).  This algorithm does not include a number of
observational effects such as noise and so is somewhat idealized;
however a comparison with a more realistic method indicated that for
high signal-to-noise spectra the differences are not large
(\cite{bry99}); we will discuss this point in more detail below.

\section{The density-temperature relation}
\label{sec:nhb}

\subsection{Testing with simulations}

\begin{figure}
\epsfxsize=3.5in
\centerline{\epsfbox{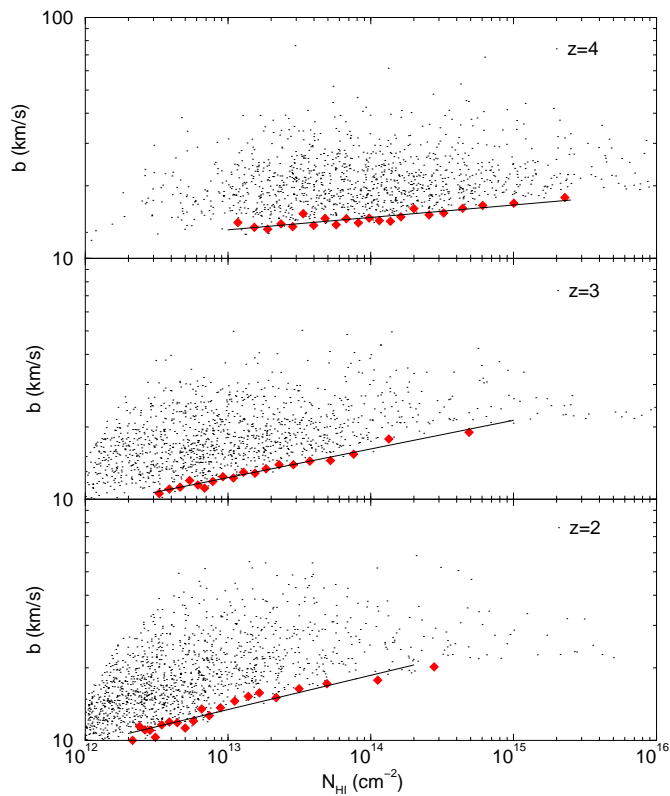}}
\caption{The column density-width ($\nh-b$) distribution
at three redshifts ($z=4,3,2$) for our LCDM model with
$\sigma_8=1.0$ and our standard heating rates.  The points
are about 1200 simulated lines, the filled diamonds trace the
minimum in the $\nh-b$ distribution as described in the text, while
the solid line is a fit to these points. }
\label{fig:nhb_he1_sig1.0_z234}
\end{figure}

In section~\ref{sec:eos}, we motivated why there should be a minimum in
the $\nh$-$b$ bivariate distribution.  In
Figure~\ref{fig:nhb_he1_sig1.0_z234}, we show this distribution for our
LCDM run with $\sigma_8 = 1.0$ and the usual Haardt \& Madau
(1996)\markcite{haa96} photo-heating rates for three redshifts: $z=4,
3$ and 2.  There is a sharp cutoff at low column densities and large
$b$ values (i.e in the upper-left corner of each frame) which is due
solely to our criterion for identifying lines, namely that the optical
depth at the line center be larger than 0.05.  More interestingly,
there is another fairly sharp cutoff at low $b$ which is the subject of
this paper.

The sharp edge that defines the cutoff is fairly obvious to the human
eye and has been previously noticed in both observations
(e.g. \cite{kir97}) and simulations (\cite{zha97a}).  In order to be
more quantitative about the position of the cutoff, we take as our
inspiration edge-detection techniques from machine-vision research.
An edge --- in one dimension --- is defined as a zero in the second
derivative of the intensity (since this is an extrema in the first
derivative, the rationale is obvious).  The application of this idea is
quite straightforward.

First, we sort the lines by column density and divide them into groups
of size 30-50 (each group is equivalent to a scan line in an image).
The smoothed density of lines is then computed as a function of $b$
with a weighted sum over all lines in each group:
\begin{equation}
\rho_b(b) = \sum_i \exp(-(b_i-b)^2/2\sigma_b^2),
\end{equation}
where $\sigma_b = 3$ km/s is the smoothing constant.  The method is
not sensitive to small changes in either this parameter or to the
number of points in the group (more lines per group mean less noise
but lower resolution along the $\nh$ direction).  We can compute
derivatives of $\rho_b$ very easily, so we simply define, for each
group with average column density $N_{HI,min}$, the edge to be at
$b_{min}$ such that
\begin{equation}
\frac{d^2\rho_b}{db^2}(b_{min}) = 0.
\end{equation}
For noisy data there are occasionally several zero crossings --- we
take the strongest, defined as the one with the largest first
derivative.  In order to get the lower cutoff, we insist that the first
derivative be position.
Software to perform this is available from the 
authors on request.

One important advantage of this algorithm is that it is relatively
insensitive to noise which tends to smear out the data (but not shift
the edge) or outliers in the $\nh-b$ distribution (which may be
rogue metal lines or simply the result of blending).  It also
non-parametric in that it doesn't assume a form for the $\nh-b$ line.

This results in a set of points which define the $\nh-b$ minimum and
are plotted in Figure~\ref{fig:nhb_he1_sig1.0_z234} as solid diamonds.
We also show a least-squares power-law fit to the line.  We used only
absorption lines within a range of column densities which was selected
to include the majority of lines but not go below about $2 \times
10^{12}$ cm$^{-2}$ or above a few times $10^{15}$ cm$^{-2}$.  The
lower limit is slightly below present day observational limits and the
upper limit marks the point were line dynamics becomes more
complicated (i.e. affected by shocks and, in the real world, star
formation).

\begin{figure}
\epsfxsize=3.5in
\centerline{\epsfbox{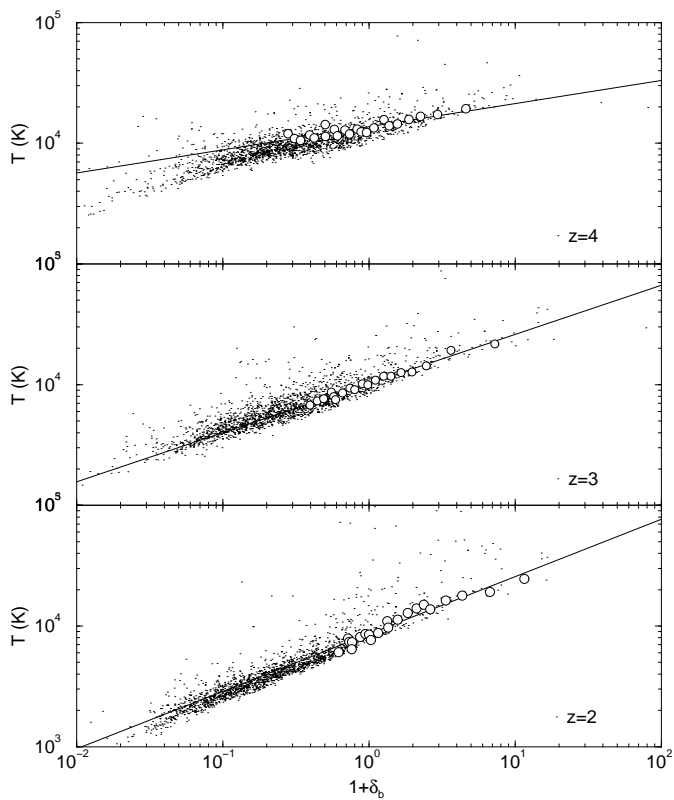}}
\caption{The temperature-overdensity relation for the same canonical
LCDM model ($\sigma_8=1.0$) in Figure~\ref{fig:nhb_he1_sig1.0_z234},
shown at three redshifts: $z=4, 3$ and 2.  The dots are 10000 random
cells, while the open circles and solid line are derived from the
minimum $\nh-b$ line as described in the text.}
\label{fig:eos_he1_z234}
\end{figure}

Each of these points may be converted into a measurement of $T$ and
$\delta_b$ via equations~(\ref{eq:nh_delta}) and (\ref{eq:bmin}).  
The results are plotted in
Figure~\ref{fig:eos_he1_z234} as open circles.  Also shown is a
measure of the temperature-density relation in the simulation by
plotting 10000 random cells as dots.  The solid line comes from
converting the power-law fits from
Figure~\ref{fig:nhb_he1_sig1.0_z234} into measurements of $T_0$ and
$\gamma$ via equation~(\ref{eq:b_nh_eos}).

The match is quite good, although clearly we preferentially probe the
upper part of the $T-\delta$ relation.  This is because, to be
observed, a line must be more dense than the surrounding gas, so
the measure is insensitive to the temperature of gas between filaments
and sheets (although note that we can still probe densities
significantly lower than the cosmic mean).  On the other end, the
maximum overdensity is around 10 because of the maximum $\nh$
limit.  If the $T-\delta$ relation is not a strict power-law, as at
$z=4$, then this can result in a substantial under- or over-estimate
of the temperature at very low or very high densities.  We also note
that there are a small fraction of points in the simulation
with moderate densities but very high temperatures.  These points tend
to lie near much more massive structures and have been enveloped in
their accretion shock.  The minimum $b$ method described here is not
sensitive to these (rare) points.

We should remind the reader of two points about the normalization of
the $\delta_b-\nh$ relation, equation~(\ref{eq:b_nh_eos}).  First,
the normalization was selected to give a good match at $z=3$ ---
changing this value is roughly equivalent to shifting the open circles
horizontally (in $1+\delta_b$).  Second, the parameter $F$ shows there
is a degeneracy amoung the parameters $\Omega_b$, $h$ and
$\Gamma_{\rm HI}$ such that as long as the value of $F$ is unchanged,
these parameters can be changed without affecting the results plotted
here.  In fact, this is one reason we choose to plot $1+\db$ rather
than a physical density.  Although the individual parameters
$\Omega_b$, $h$ and $\Gamma_{\rm HI}$ are not well determined,
this particular combination is, from observations of the Ly$\alpha$
forest (see, for example, \cite{rau97}).

\subsection{Changing the equation of state}

Given our success in measuring the temperature of the IGM in the
canonical simulation described in the previous section, it is
interesting to see if we can detect the effect of changing the primary
heating mechanism as outlined in the introduction.  In this section,
we show that this is possible.  We retain the same cosmological model,
but modify the HeII photo-heating rate as described in Abel \&
Haehnelt (1999)\markcite{abe99}.  This takes into account radiative
transfer effects during HeII photo-ionization that are neglected in
these simulations.  Since the amplitude of the effect is hard to
gauge, we either multiply the rate by two or four.

\begin{figure}
\epsfxsize=3.5in
\centerline{\epsfbox{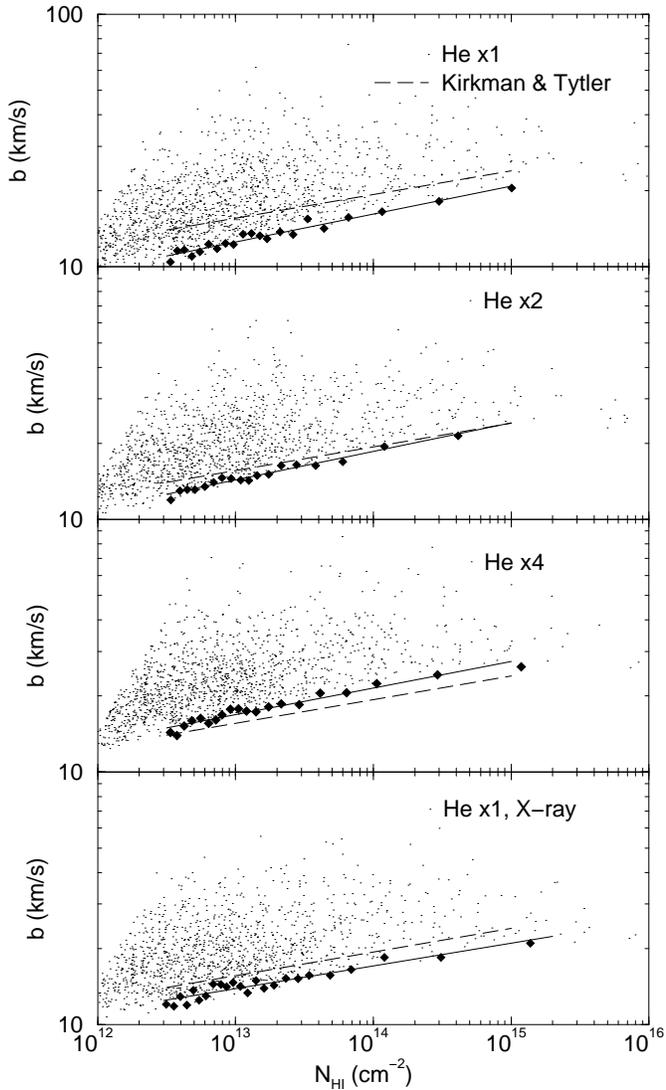}}
\caption{The column density-width ($\nh-b$) distribution
at $z=2.7$ for our canonical LCDM model with four different heating
rates.  The top frame shows the usual Haardt \& Madau (1996) HeII
photo-heating rate, while the next two panels demonstrate the effect of
increasing this rate by factors of two and four, respectively.  The
bottom panel has the usual HeII rate but includes Compton X-ray
heating.  The filled diamonds describe the minimum in the $\nh-b$
distribution as described in the text.  The solid line is a fit to
these points, and the dashed line is the same for each panel and shows
the observational determination of the $\nh-b$ minimum from Kirkman \&
Tytler (1997). }
\label{fig:nhb_eos_sig1.0}
\end{figure}

\begin{figure*}
\epsfxsize=6.0in
\centerline{\epsfbox{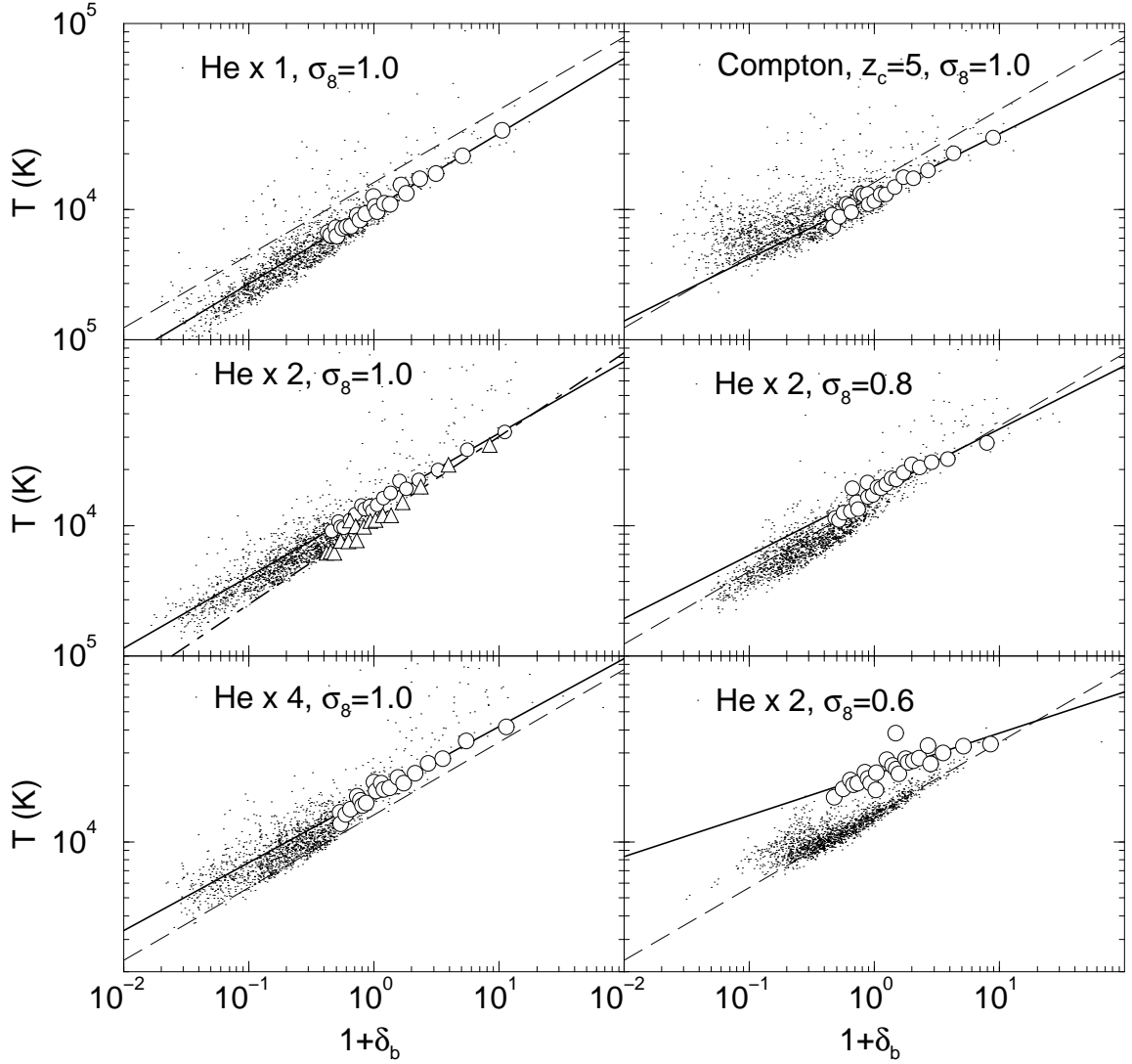}}
\caption{The temperature-overdensity relation at $z=2.7$ for six
simulations.  The symbols are as in Figure~\ref{fig:eos_he1_z234}.
The three left panels are from the canonical LCDM model (with
$\sigma_8=1.0$) for the same three values of HeII photo-heating rate
as in Figure~\ref{fig:nhb_eos_sig1.0}.  The upper right panel is for
the same model but with X-ray Compton heating instead.  The two lower
panels on the right were run with twice the usual Haardt \& Madau HeII
photo-heating rate but for low-power LCDM models, with $\sigma_8=0.8$
and $\sigma_8=0.6$.  For comparison, each plot also shows the equation
of state from the middle left frame as a dashed line.  The triangles
and dot-dashed line in the middle-left panel show the equation of
state derived from a different Voigt-profile fitting technique as
described in the text.}
\label{fig:eos_z3}
\end{figure*}

The results are shown in Figure~\ref{fig:nhb_eos_sig1.0}, again with
$b_{min}$ points and fit as determined from our edge-detection
algorithm.  Here, in order to give a concrete comparison with
observations and to provide a constant reference point, we also plot
the observational equivalent of the $\nh-b$ minimum as found by
Kirkman \& Tytler (1997)\markcite{kir97} at a mean redshift of 2.7 for
a single line-of-sight.  Although they fit this by eye the result
agrees very well with the method used here.  Our standard HeII
photo-heating rate produces temperatures which are too low, while the
x2 and x4 simulations are much closer and bracket the result, with the
x2 case the closest.  There is some evidence that the slope is in
disagreement for all cases; however, this does not appear to be
strongly significant given our level of uncertainty (see below).  We
can apply the same method used earlier to determine the $T-\delta$
relation, which is shown in the three left panels of
Figure~\ref{fig:eos_z3}.  Again, the equation of state is quite
accurately determined.

Next, we examine the importance of Compton X-ray heating in the bottom
panel of Figure~\ref{fig:nhb_eos_sig1.0}, which is again our canonical
LCDM simulation with the usual HeII photo-heating rates.  However, now
we include hard X-ray heating as described earlier.  While this does
boost the temperature somewhat, it is clearly --- by itself ---
insufficient to match observations.  Since the Compton heating rate is
independent of density, it tends to flatten the temperature-density
relation, shown in the upper right panel of Figure~\ref{fig:eos_z3}.
However, the effect at $z=2.7$ is mostly limited to low densities and
so is very difficult to detect with the minimum $\nh-b$ method.

\subsection{Changing the power spectrum}

Although we have been successful in measuring the equation of state
for our canonical LCDM model, we argued in section~\ref{sec:theory}
that the amplitude of fluctuations on scales of a few hundred kpc is
also important in determining the distribution of line widths.  In
this section, we will demonstrate that for some models, this effect
prevents us from accurately measuring the temperature-density
relation.

\begin{figure}
\epsfxsize=3.5in
\centerline{\epsfbox{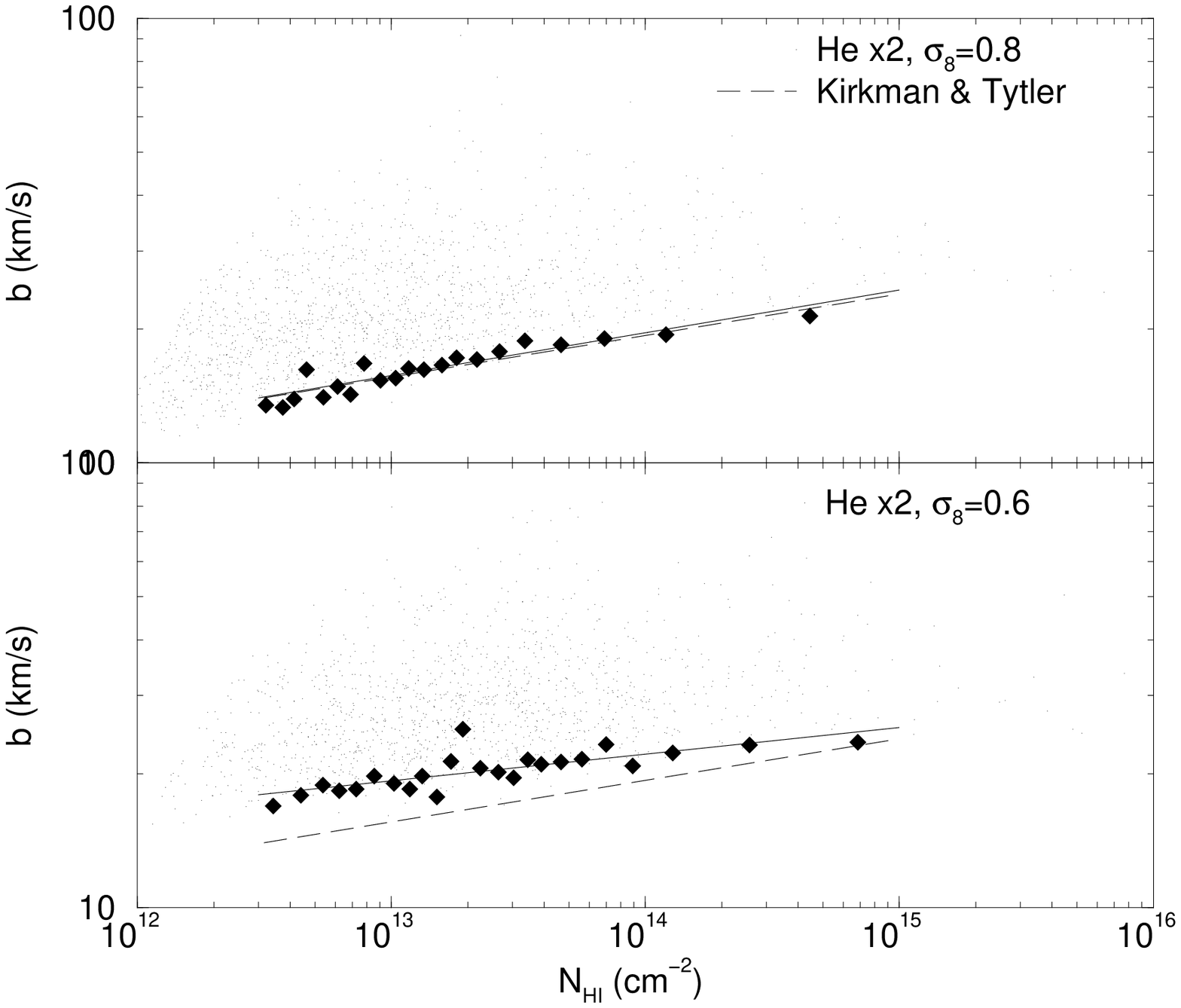}}
\caption{The column density-width distribution ($\nh-b$) at $z=2.7$
for two LCDM simulations.  These both use twice the usual HeII heating
rate, but decrease the amplitude of the initial fluctuation spectrum
to $\sigma_8=0.8$ (top) and $\sigma_8=0.6$ (bottom).  The symbols are
as in Figure~\ref{fig:nhb_eos_sig1.0}.}
\label{fig:nhb_power}
\end{figure}

Figure~\ref{fig:nhb_power} shows the results from two LCDM models in
which the power has been reduced to $\sigma_8 = 0.8$ and
$\sigma_8=0.6$.  This changes the derived minimum $\nh-b$ line.  For
$\sigma_8=0.8$ (the top panel) the match with observations is very
good, while the lower power simulation produces a power-law fit that
is too flat.  Since the equation of state has not changed, application
of the $\nh-b$ minimum method results in temperatures which are too
hot, particularly for the lowest-power run.  This shown in the bottom
right panels of Figure~\ref{fig:eos_z3}.

This happens because one of our key assumptions is violated:
specifically that there be a substantial number of lines for which
$b_{vel}$ be small.  For the lower-power models, the smallest
filaments are almost all in the pre-turnaround stage.  That is, their
peculiar velocities are still smaller than the Hubble flow across
their width, so that these two values cannot cancel.  This
preferentially affects low column density lines because they are the
smallest fluctuations.  In fact, lines with a column density of around
$\nh \sim 10^{15}$ cm$^{-2}$ ($1+\db \sim 10$) faithfully reproduce
the correct temperature even for the $\sigma_8=0.6$ simulation.

This demonstrates that the minimum $\nh-b$ method suffers from a
degeneracy between the gas temperature and the amplitude of
fluctuations.  As long as $\sigma_{34} \gtrsim 1.6$, the method can be
used in a straightforward fashion (we use $\sigma_{34}$ since this is
much closer to the scale and redshift of interest and so is nearly
independent of other cosmological parameters).  Below this value, there
is still useful information to be gained, but the interpretation is
more complicated.  In particular, without other knowledge about
$\sigma_{34}$, the value of $T_0$ derived in this way is an upper limit
and the value of the slope $\gamma$ is a lower limit.

Although we do not show the results here, we have also analyzed an
LCDM simulation with a lower value of the Hubble constant as well as
an SCDM model (see Table 1).  The results agree with the trends
discussed in this section.

\begin{figure}
\epsfxsize=3.5in
\centerline{\epsfbox{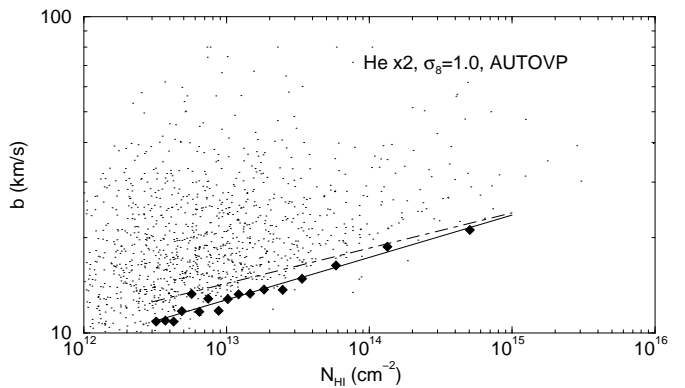}}
\caption{The column density-width distribution ($\nh-b$) at $z=2.7$
for a LCDM simulation with $\sigma_8=1.0$ and twice the usual HeII
photo-heating rate.  This is the same simulation analyzed in the
second panel of Figure~\ref{fig:nhb_eos_sig1.0}, but here we use
a different Voigt-profile fitting technique.  The diamonds and solid
line are the derived minimum $\nh-b$ line, while the dot-dashed line
is the minimum $\nh-b$ fit found previously, for comparison.
}
\label{fig:nhb_autovp}
\end{figure}

It is important to ask at this point what the uncertainties are in
determining the minimum $\nh-b$ line.  The primary source of
uncertainty is fitting the Voigt-profiles in the first place, as this
is both non-linear and non-unique (e.g. \cite{kir97}).  In order to
gauge the magnitude of the possible error, we select one simulation
(LCDM $\sigma_8=1.0$ with twice the usual HeII photo-ionizing rate)
and fit it with the more realistic method AUTOVP (\cite{dav97}),
kindly provided by Romeel Dav\'e.  This method performs a chi-squared
minimization to produce the line list from the simulated spectrum.
Figure~\ref{fig:nhb_autovp} shows the result for a signal-to-noise
ratio of 60, along with the fit found with the more idealized
Voigt-profile fitting algorithm.  Clearly there is some difference,
which also affects the derived equation of state, shown in the
middle-left panel of Figure~\ref{fig:eos_z3}.  This amounts to about a
15\% difference in $T$ at $\db=0$, and is mostly due to fitting
Voigt-profiles to lines which do not follow this profile in detail.

\section{The median of the $b$-distribution}
\label{sec:median}

\begin{figure}
\epsfxsize=3.5in
\centerline{\epsfbox{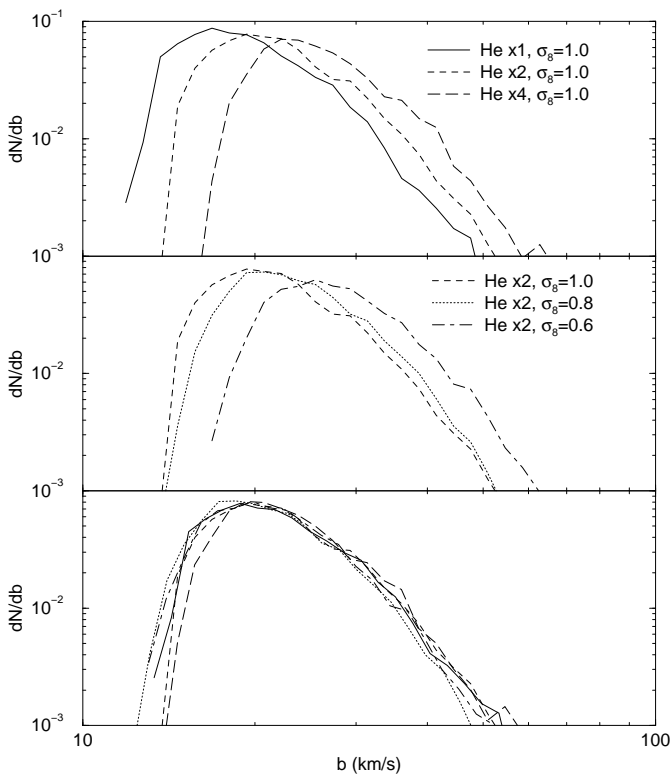}}
\caption{The simulated $b$ distribution function for three
models with the same power but different temperatures (top),
the same temperature but differing power (middle) and all
models scaled as described in the text(bottom).}
\label{fig:dndb_all}
\end{figure}

We have so far focused on the low $b$ cutoff, but now turn, briefly,
to the rest of the distribution.  The top panel of
Figure~\ref{fig:dndb_all} shows $dn/db$ at $z=2.7$ for our three
$\sigma_8=1.0$ models with varying equations of state (i.e. different
HeII photo-heating rates).  Only lines in the range $\nh =
10^{13.1}-10^{14}$ cm$^{-2}$ are used so as not to be biased by the
line-selection function.  All distributions are normalized so that
$\int(dn/db)db = 1$.  This plot shows that increasing the temperature
results in a constant shift in $\log(b)$.  As discussed in Bryan \etal
(1999), this is not primarily a result of thermal Doppler broadening,
but comes instead from a thickening of the filaments and sheets (in
both physical and velocity space) due to the influence of the
increased pressure --- gas is driven out of the centers of the
filaments.

\begin{figure}
\epsfxsize=3.5in
\centerline{\epsfbox{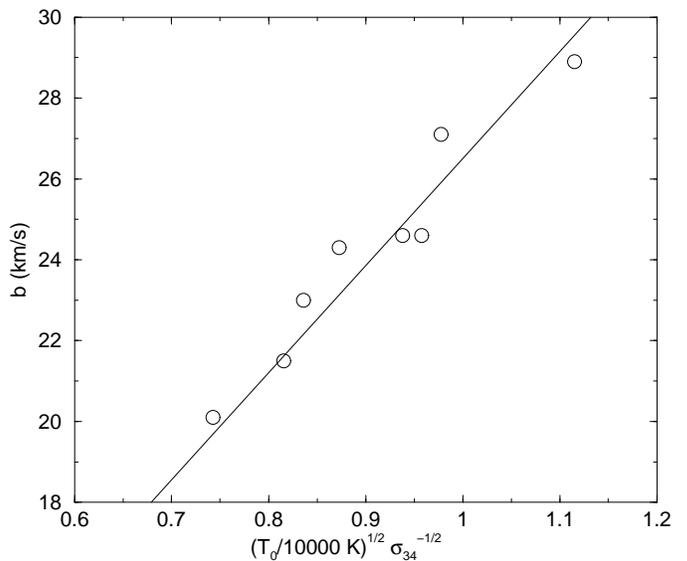}}
\caption{The median of the $b$ distribution depends almost
entirely on just two parameters: the temperature of the gas
($T_0^{\prime}$) and the amplitude of the power spectrum
($\sigma_{34}$).}
\label{fig:bmed}
\end{figure}

The middle panel of Figure~\ref{fig:dndb_all} shows the effect of
changing the amplitude of the power spectrum while keeping the
temperature constant.  The results appear to match the simple scaling
derived in section~\ref{sec:theory}.  This degeneracy between
temperature and power can be written in terms of the median of this
distribution:
\begin{equation}
\label{eq:bmed}
b_{med}(T_0^\prime, \sigma_{34}) = 
      26.5 {\rm km/s} \left( \frac{T_0^{\prime}}{10000 K} \right)^{1/2}
         \left( \sigma_{34} \right)^{-1/2}
\end{equation}
We use $T_0^\prime$ to indicate the temperature at $\delta_b=0$
measured directly from the simulations, rather than via the minimum
$\nh-b$ method.  In Figure~\ref{fig:bmed}, we plot this function
against the measured median $b$ parameters from our simulations. 
Finally, in order to demonstrate that there is not much change in the
shape, we rescale the $b$ distributions in the top two panels of
Figure~\ref{fig:dndb_all} with the following transformation: $b
\rightarrow fb$ and $(dn/db) \rightarrow (dn/db)/f$.  We define $f =
b_{med}(T_0^\prime,\sigma_{34})/b_{med}(13000 K, 1.93)$, where
$b_{med}$ is given in equation~(\ref{eq:bmed}).  The result is shown
in the bottom panel of this figure.  The scaling works very well,
indicating that the shape of the distribution changes little, once the
median is specified.  The biggest differences are at small $b$, where
thermal Doppler broadening dominates.

Our simulated boxes are too small to fully contain all the large-scale
power.  Previously (\cite{bry99}) we showed that this has little
effect on the shape of the $b$ distribution but can cause a small
shift in the median.  In order to gauge the size of this effect here,
we ran two models with $\sigma_8=0.8$, one with our usual
box-length of 4.8 Mpc and one with twice this size (see
Table~\ref{table:simulations}).  The both had the same cell size
(i.e. the larger box simulation had $256^3$ cells rather than our
more usual $128^3$).  
The change in the median was only 0.4 km/s (1.6\%).  This could be
slightly larger for models with more power, but is unlikely to be a
significant effect.

\section{A preliminary comparison to observations}
\label{sec:observations}

\begin{figure}
\epsfxsize=3.5in
\centerline{\epsfbox{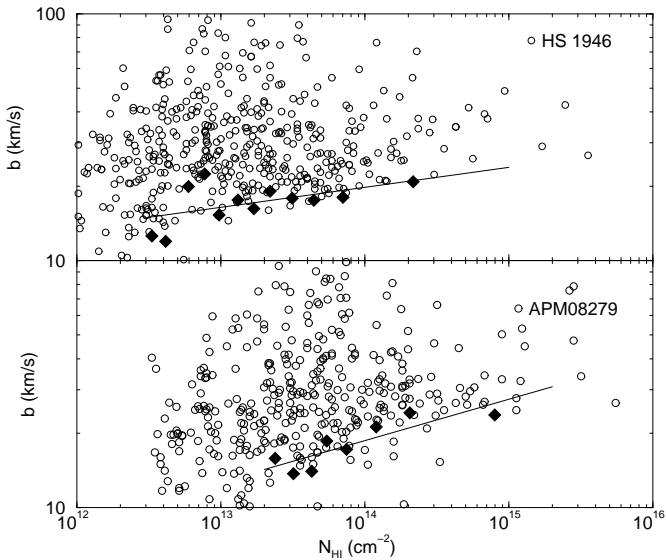}}
\caption{The column density-width distribution ($\nh-b$) at for two
sets of observed absorption line systems.  The top is HS 1946+7658
(\cite{kir97}) at $<z>=2.7$ and the bottom is from APM 08279+5255
(\cite{ell99}) at $<z>=3.4$.  The symbols are as in
Figure~\ref{fig:nhb_eos_sig1.0}.}
\label{fig:nhb_obs}
\end{figure}

In this section, we make a preliminary comparison to observations,
using previously published results from the quasar HS1946+7658
(\cite{kir97}) at a mean redshift $<z> = 2.7$ and APM 08279+5255
(\cite{ell99}) at $<z>=3.4$.  Both observations have high
signal-to-noise ratios, ranging from 15 to 100 for HS1946+7658 and 30
to 150 for APM 08279+5255.  It should be kept in mind that the Voigt
profile fitting technique used in these papers (which is not fully
automated) differs somewhat from both methods used here.  The $\nh-b$
distributions are shown in Figure~\ref{fig:nhb_obs}, along with the
derived minimums using our edge-detection method.  (for APM 08279+5255
we adopt a minimum column density of $2 \times 10^{13}$ cm$^{-2}$ due
to concerns about line-blending at these high redshifts).

\begin{figure}
\epsfxsize=3.5in
\centerline{\epsfbox{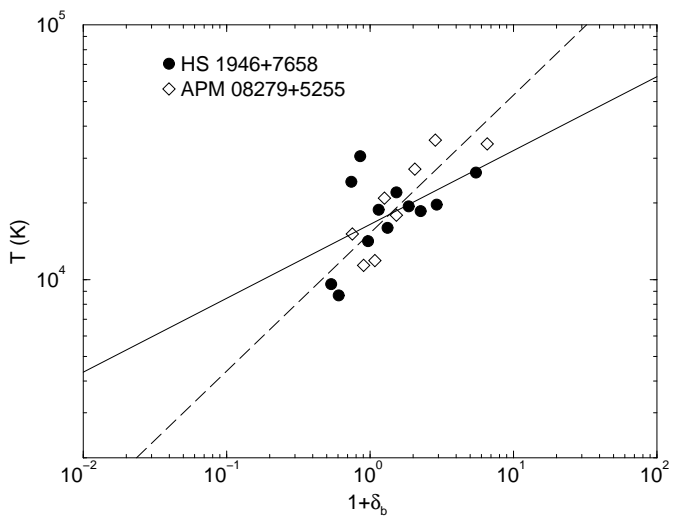}}
\caption{The temperature-density relation as derived from
the two quasars analyzed in Figure~\ref{fig:nhb_obs}.} 
\label{fig:eos_obs}
\end{figure}

The minimum $\nh-b$ edges detected can be converted into a measurement
of the temperature-density relation, which is shown in
Figure~\ref{fig:eos_obs}.  The values of $F$ required to convert
column density to $1+\delta_b$ were determined by fitting the column
density distribution to the simulations (we use the LCDM
$\sigma_8=0.8$ simulation but this is not very sensitive to which
model we select).  The noise in the $T-\delta_b$ relation is larger
than for the simulations because of the much smaller number of lines
($\sim 300$ as compared to $\sim 1200$).
The power law fits are
given by $T_{0,4} = 1.65$, $\gamma = 1.29$ (HS1946+7658) and $T_{0,4}
= 1.52$, $\gamma=1.54$ (APM 08279+5255).  We remind that reader that
these determinations are really upper limits to the temperature rather
than measurements because of the possible effects of cosmology
(i.e. the unknown value of $\sigma_{34}$).

There is an indication from the higher-redshift system that the gas is
cooler at low density (i.e. a steeper equation of state), although
clearly this is substantialy uncertainty.
Still, a similar trend of lower $b$ lines at higher redshift has been
previously noted from different data (\cite{hu95}; \cite{kim97}), so
it is worth considering the possibility that (low density) gas is
cooler at higher redshift.  This does not agree with what is expected
for gas dominated by steady photo-ionization heating and adiabatic
cooling: $T_0 \sim (1+z)^{1.7}$ with a slowly steepening equation of
state slope (\cite{mei93}; \cite{mir94}; \cite{hui97b}; \cite{abe99}).
An alternate heating source, such as late Helium reionization, would
be required if this result proves true.

\begin{figure}
\epsfxsize=3.5in
\centerline{\epsfbox{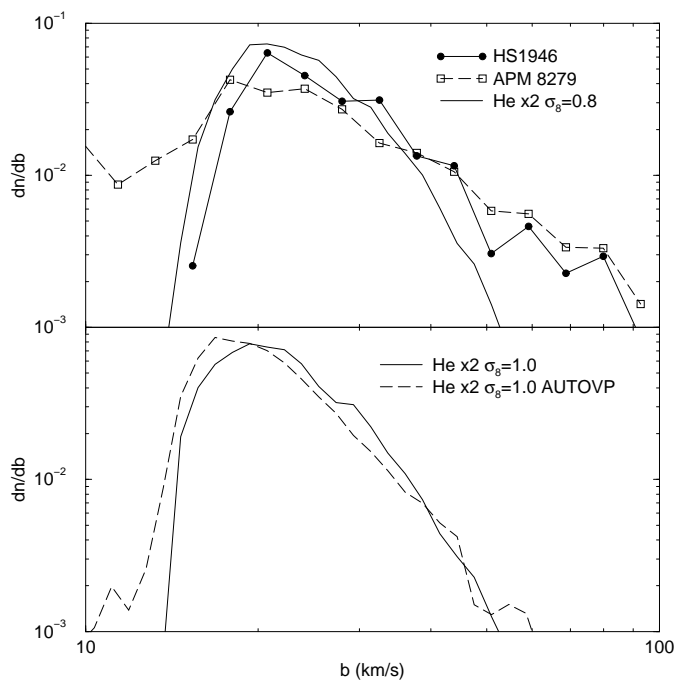}}
\caption{Top: $b$ distributions from two observed quasar line systems
along with one of the better fitting simulated results; bottom:
$dn/db$ for the same simulation using two different Voigt-profile
fitting algorithms: the solid line shows ZANM97 (\cite{zha97b}) which
is quite idealized, while the dashed line indicates AUTOVP
(\cite{dav97}) which includes random noise with a signal-to-noise 
ratio of 60.}
\label{fig:dndb_obs}
\end{figure}

To compare the shape of the $b$ distributions to observations, in
Figure~\ref{fig:dndb_obs} we plot $dn/db$ from the same two quasar
systems previously discussed.  We also show the distribution from our
LCDM $\sigma_8=0.8$ simulation with twice the Haardt \& Madau HeII
photo-heating rate.  The shape is in reasonable agreement with
HS1946+7658, but the other system has a large number of very low $b$
lines which are not seen in any of the models considered here
(although a very low temperature model might match).  Both observed
systems also have a more pronounced non-Gaussian tail at large $b$ 
than appears in the simulations.
Note, however, that the log-log plot accentuates this tail when
compared to the more usual linear plot.  It should also be kept in
mind that the Voigt-profile fitting method used for the observations
differs from either employed in this paper.  Clearly, a more
definitive result will require identical treatment of data and
simulations.  The bottom panel of the same Figure shows how the two
different Voigt-profile fitting algorithms used in this work compare.

The observed median $b$ (in the same column density range considered
earlier) is 27.3 for both systems.  Using equation~(\ref{eq:bmed}),
this implies that
\begin{equation}
\left( {T_0^{\prime}}/{10000 K} \right)^{1/2} \left(
\sigma_{34} \right)^{-1/2} = 1.03.
\end{equation}
If we use the upper limits for $T_0$ derived earlier from the $\nh-b$
distribution, then we can get an upper limit on the amplitude of the
power spectrum: $\sigma_{34} \lesssim 1.52$.  The uncertainty is about
20\%, due to the uncertainty in the measured temperature.  This is
quite close to the minimum value of $\sigma_{34}$ required for a
straightforward interpretation of the $\nh-b$ minimum method
($\sigma_{34} \gtrsim 1.6$).  Interestingly, this value of
$\sigma_{34}$ (around 1.5-1.6) agrees well with a number of
determinations of the power spectrum amplitude using other
characteristics of the Ly$\alpha$ forest (\cite{gne98};
\cite{cro99b}).  For the LCDM model it is also in accordance with the
normalization from COBE and rich clusters of galaxies
(e.g. \cite{lid96}).

\section{Conclusions}

In this paper, we have investigated a method for determining the
relation between density and temperature (loosely denoted the equation
of state of the IGM) from the distribution of quasar absorption lines
in the $\nh-b$ plane.  Specifically, we look for a sharp minimum
line in this plane arising from the fact that Doppler thermal
broadening sets a minimum line width.  Because there is a tight
relation between column density and overdensity, we can relate
$\nh$ to overdensity and $b$ to temperature.  We derive a simple
model which reproduces this behaviour, and state clearly its
assumptions.

We test this method with a range of equations of state, including an
enhanced HeII photo-heating rate (assumed to be due to neglected
optical transfer effects) and X-ray Compton heating.  We show that the
method works as long as the power spectrum amplitude is sufficiently
high, so that $\sigma_{34} \gtrsim 1.6$.  If the density fluctuations
are too small, then one important assumption fails: that there be a
substantial fraction of lines whose width is dominated by thermal
broadening.  When this occurs, it mimics an equation of state which is
hotter and flatter.


Very recently, Schaye \etal (1999)\markcite{sch99} independently
investigated the feasibility of using this method.  They used a
different method for identifying the $\nh-b$ cutoff, but came to quite
similar conclusions as presented here, with one important exception.
They argued that there was no cosmological dependence on the cutoff in
the $\nh-b$ plane, in disagreement with the results in this paper.
However, they examined a relatively small number of simulations which
all had similar values of $\sigma_{34}$, mostly comparable to, or
somewhat larger than, the critical value listed above.  In this case,
it would be very difficult to notice the effect due to power.

Applying our results to two quasar lines-of-sight with mean redshifts
of $z=2.7$ and $z=3.4$, we derive a temperature-density relation from
these two systems that is similar to those found in at least some of
the simulations presented here.  We find a temperature of
approximately $16000$ K for gas with the mean density rising to about
$35000$ K for an overdensity of six.  The uncertainty of these numbers
is around 20\%, however without any more information about the value
of $\sigma_{34}$, they must be treated as upper limits to the
temperature of the gas.  If we were to assume that the power spectrum
criterion is satisfied then this represents fairly hot gas compared to
traditional models.  The additional HeII photo-heating is sufficient
to produce this much heat; however, by itself Compton X-ray heating is
not.

We turn now from the $\nh-b$ minimum to $dn/db$, the distribution of
$b$ parameters.  Following similar earlier work (\cite{hui98}), we
present a simple linear argument which shows that the other important
parameter in determining the $b$ parameter distribution is the
amplitude of the power spectrum.  We demonstrate that simulations
reproduce this scaling ($b_{med} \sim T^{1/2}\sigma_{34}^{-1/2}$), and
show that the shape of the $b$ distribution stays nearly invariant to
changes in temperature or the power spectrum amplitude.  Its median
value can be given as a simple function of the gas temperature and
$\sigma_{34}$ (at $z\sim 3$).  If we use the temperature derived from
the $\nh-b$ method, this implies that $\sigma_{34} \sim 1.5$ (with an
uncertainty of about 20\%), a value which is in reasonable agreement
with other methods of determining the amplitude of the power spectrum.
It should be kept in mind that since the temperature measurement is
really an upper limit, then this value for $\sigma_{34}$ is also an
upper limit.

Theuns \etal (1999)\markcite{the99} also investigated the cosmological
dependence of the $b$-distribution.  They found the results sensitive
to the gas temperature (as we have here), but did not find the
sensitivity to power spectrum discussed here.  Again, it seems likely
that this is due to the small range in spectral amplitudes of their
simulations.

The degeneracy described in this paper between power and temperature
means that the $b$ distribution alone will not be sufficient to
determine the equation of state of the gas.  This is unfortunate
because the evolution of the temperature-density relationship can
provide constraints on other cosmological interesting events.  For
example, if the gas were to be colder above $z=3$ (as the data
presented here might be indicating), one possible explanation would be
the late reionization of helium (\cite{rei97}; \cite{hae97};
\cite{abe99}).  This degeneracy can be broken by using other aspects
of the Ly$\alpha$ forest (e.g. \cite{cro98}; \cite{mac99}) to
independently fix the amplitude of fluctuations at these scales and
redshifts.

\vspace{0.2cm} 

We acknowledge useful discussion with Tom Abel, Piero Madau, Avery
Meiksin and Lam Hui.  Some of data presented herein were obtained at
the W.M. Keck Observatory, which is operated as a scientific
partnership amoung the California Institute of Technology, the
University of California and the National Aeronautics and Space
Administration.  The Observatory was made possible by the generous
financial support of the W.M. Keck Foundation.  This work is done
under the auspices of the Grand Challenge Cosmology Consortium and
supported in part by NSF grants ASC-9318185 and NASA Astrophysics
Theory Program grant NAG5-3923.  Support for this work was also
provided by NASA through Hubble Fellowship grant HF-01104.01-98A from
the Space Telescope Science Institute, which is operated by the
Association of Universities for Research in Astronomy, Inc., under
NASA contract NAS6-26555.

\end{document}